# Accurate atomic positions via local-orbital tomography with depth-dependent interactions

**Authors**: Liangze Mao[1,2,3†], Jizhe Cui[1,2,3†] and Rong Yu[1,2,3*]

**Affiliations**:

[1]School of Materials Science and Engineering, Tsinghua University, Beijing 100084, China.

[2]MOE Key Laboratory of Advanced Materials, Tsinghua University, Beijing 100084, China.

[3]State Key Laboratory of New Ceramics and Fine Processing, Tsinghua University, Beijing 100084, China.

*Corresponding author. Email: ryu@tsinghua.edu.cn

†These authors contributed equally to this work.

## Abstract

**Three-dimensional reconstruction of atomic structure, known as atomic electron tomography (AET), has found increasing applications in materials science. The AET has been limited to very small nanoparticles due to the challenges of obtaining accurate atomic positions for large objects, for which the projection approximation generally assumed in AET is no longer valid due to the evolution of the electron probe along the beam direction. To address these challenges, we have developed a method that represents atoms as local orbitals and implements a depth-dependent probe-object interaction. Compared to conventional tomographic methods, the accuracy of atomic positions is improved by a factor of three and provides a solution for the tomographic reconstruction of large objects.**

# Introduction

Electron tomography allows the determination of the three-dimensional (3D) structure of an object from its projections [1-5]. Recently, atomic resolution has been achieved in electron tomography, leading to extensive analysis of atomic structure of crystalline and amorphous alloys, mostly in nanoparticle form [6-12]. The generally used tomographic algorithms such as weighted back projection (WBP) [13], algebraic reconstruction technique (ART) [14], simultaneous algebraic reconstruction technique (SART) [15], simultaneous iterative reconstruction technique (SIRT) [16], equal slope tomography (EST) [17], generalized Fourier iterative reconstruction (GENFIRE) [18], and real space iterative reconstruction (RESIRE) [19], have been developed to achieve high-quality 3D reconstruction results.

These tomographic algorithms describe an object as a 3D matrix consisting of voxels [13-19]. The reconstructed 3D matrix's local maxima are considered potential atoms, and their coordinates are determined through a post-processing procedure, specifically peak finding [9, 20]. The accuracy of atomic positions can be substantially reduced due to the 'missing wedge' problem [21, 22], resulting in elongated and blurred voxels, which significantly reduces the accuracy of the atom positions [23, 24]. To address this issue, recent developments have introduced neural network techniques [25, 26]. Additionally, the positional accuracy of the voxel-based reconstruction algorithms is limited by the voxel size. Reducing the voxel size to improve accuracy would increase the computational cost, thereby limiting the positional accuracy.

In addition, typical tomography reconstruction algorithms generally require the use of the linear projection approximation [2, 27-31]. The 2D image of an object in a given orientation is evaluated by integrating the object in the direction of the electron beam. This assumption is acceptable for ultrathin samples (<~8 nm) when the depth of focus is much larger than the object. The electron probe evolves along the beam direction, and its interaction with the object varies with the depth of the probe. Neglecting this depth-dependent interaction reduces positional accuracy for larger objects and limits AET reconstruction to very small objects.

This study introduces a new tomography reconstruction algorithm named local-orbital tomography (LOT). Unlike conventional AET algorithms, LOT represents an object as a sum of local-orbital functions instead of a 3D matrix. Additionally, the depth-dependent probe-object interaction is handled by dividing both the probe and object into multiple slices and allowing the probe to evolve in the depth direction. The positional accuracy of atoms can be achieved at a high level using the LOT algorithm, even in large-size samples.

# Local-orbital tomography (LOT)

Considering the atomic nature of matter, we represent a 3D object as a sum of local-orbital functions, rather than a 3D matrix. The 3D object is divided into multiple equally spaced slices along the beam direction. The projection of the *k*-th slice at the *j*-th angle is written as:

$$O_{jk}(u,v) = \sum_{l=1}^{N_{jk}} D_{ij}(u,v) \tag{1}$$

where $N_{jk}$ is the number of local orbital functions within the *k*-th slice at the *j*-th angle.

Each local-orbital function represents an atom. The form of the local-orbital function can take the form of a 3D Gaussian function, Slater function, numerical atomic orbital, or other local functions. For instance, the projection of a local-orbital function can be expressed as a 3D Gaussian function:

$$D_{ij}(u,v) = H_i e^{\frac{[u-(u_{ij}+u_j)]^2 + [v-(v_{ij}+v_j)]^2}{B_i}} \tag{2}$$

where $D_{ij}$ represents the 2D projection of the *i*-th localized orbital function at the position (*u,v,w*) and the *j*-th angle, $H_i$ and $B_i$ represent the intensity and width of the *i*-th localized orbital function, respectively. ($u_{ij}$, $v_{ij}$, $w_{ij}$) are the central 3D position coordinates of the *i*-th localized orbital function at the *j*-th angle. ($u_j$, $v_j$, $w_j$) are the 3D directional drift of the center of the localized orbital function in the *j*-th angle relative to the true position.

$$\begin{bmatrix} u_{ij} \\ v_{ij} \\ w_{ij} \end{bmatrix} = R(\psi_j, \theta_j, \varphi_j) \cdot \begin{bmatrix} x_i \\ y_i \\ z_i \end{bmatrix} \tag{3}$$

where the *j*-th angle is represented by the Euler angles ($\psi_j$, $\theta_j$, $\varphi_j$) corresponding to the rotation around the *x*, *y*, and *z* axes, respectively, ($x_i$, $y_i$, $z_i$) are the central 3D position coordinates at 0°.

HAADF images are commonly used to generate images at different angles due to their good linearity with the mass-thickness of an object in the probe direction. To account for depth-dependent probe-object interaction, both the probe and object are divided into multiple slices, and the probe's evolution in the depth direction is allowed [32], as shown in **Fig. 1**. The probe-object interaction is analyzed using a multilayer approach. The calculated image at a given angle is expressed as:

$$I_j^{cal}(u,v) = \sum_{k=1}^{n} p_{jk}(u,v) \otimes O_{jk}(u,v) \tag{4}$$

where $p_{jk}(u,v)$ is the probe function propagated to the *k*-th slice at *j*-th angle, $\otimes$ is the convolution symbol, *n* is the total number of slices. There are various options for the form of the probe function, including generating it using known imaging conditions such as defocus, spherical aberration, and astigmatism. If the probe functions for each slice remain unchanged during reconstruction, the probe is referred to as rigid depth-dependent interaction, or 'rigid DDI' for short. If the probe functions are permitted to optimize on reconstruction, the probe is referred to as adaptive depth-dependent interaction, or 'adaptive DDI' for short. Conventional AET algorithms assume a direct projection, which means that the probe function is independent of depth and angle, i.e., $p_{jk}(u,v) \equiv 1$.

The optimization process employs a mean square difference loss function, and the parameters within the loss function are iteratively optimized using gradient descent:

$$\varepsilon = \frac{1}{2mpq} \sum_{j=1}^{m} \sum_{u=1}^{p} \sum_{v=1}^{q} [I_j^{obs}(u,v) - I_j^{cal}(u,v)]^2 \qquad (5)$$

where $I_j^{obs}(u,v)$ and $I_j^{cal}(u,v)$ represent the *j*-th measured and calculated image with a size of $p \times q$ pixels and *m* is the number of images.

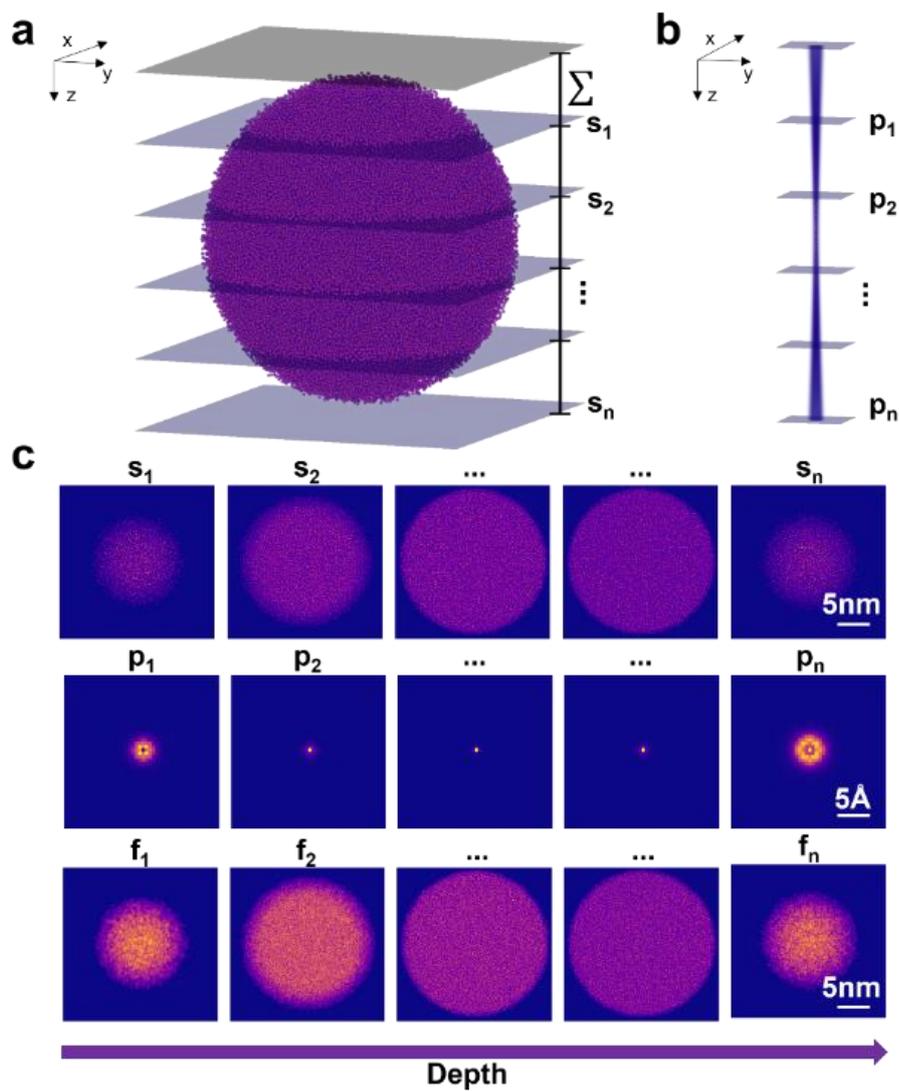

**Fig. 1. Schematic of the calculated imaging process and reconstruction methods. a**, The 3D object is divided into multiple equally spaced slices along the direction of the electron beam: $L_1$, $L_2$, ..., $L_n$ [33]. **b**, The probe in three dimensions is sliced at the corresponding depths of each slice. **c**, The convolution result of the linear projection $s_1$, $s_2$, ... $s_n$ of the 3D object in each slice and the corresponding probe slice $p_i$ at the depth of the slice is shown as $f_n$.

# LOT reconstruction on a crystalline particle

The following section presents the simulation test of the LOT method for both crystalline and amorphous particles. The atomic structure of the objects is known for simulation tests, making it convenient to compare the accuracy of the different reconstruction results. First, a Pt nanoparticle with the five-fold twin structure was constructed, composed of 25,790 Pt atoms, as shown in **Fig. 2**a. This model is referred to as $M_0$ throughout the text. Multislice simulations [34] were subsequently conducted on model $M_0$ to generate HAADF-STEM images at tilting angles ranging from -72° to 70° with an increment of 2°. The simulation parameters used were a slice thickness of 2 Å, a sampling interval of 0.1 Å for both the probe and object, zero $C_3$ and $C_5$ aberration, an acceleration voltage of 300 keV, a convergence semi-angle of 25 mrad, and inner and outer detector angles of 30 mrad and 195 mrad, respectively. To reduce the channeling effect, the model was rotated away from low-index zone axes. **Fig. 2**b shows the simulation image at 0°.

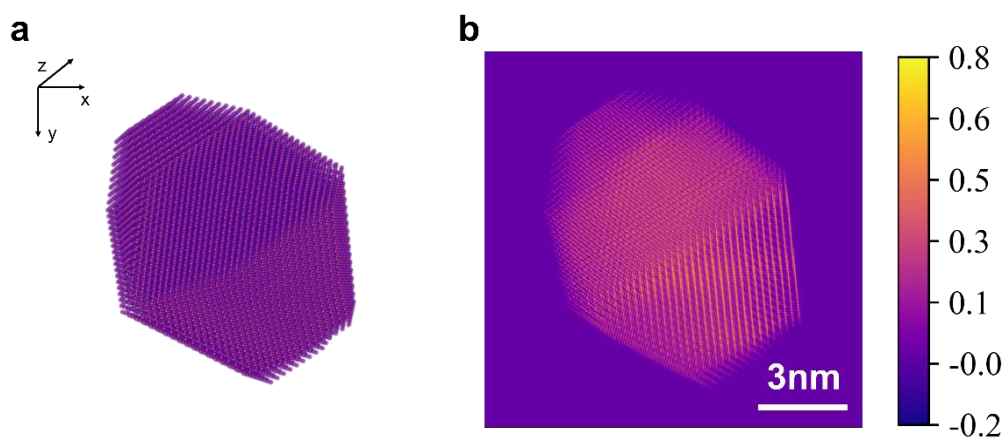

**Fig. 2. Schematic of the five-fold twin structure $M_0$. a**, The atomic model of $M_0$. **b**, The simulated HAADF-STEM image of $M_0$ at 0°. Scale bar, 3nm.

Tomographic reconstructions were performed using the simulation images and the three different forms of the probe functions. **Fig. 3** compares the reconstruction results for the different probe functions. The R-factor (see "Methods" section) denotes the difference between the reconstruction results and the simulation images. It is clear that the quality of the reconstruction improves as the probe function changes from direct projection to rigid DDI and then to adaptive DDI.

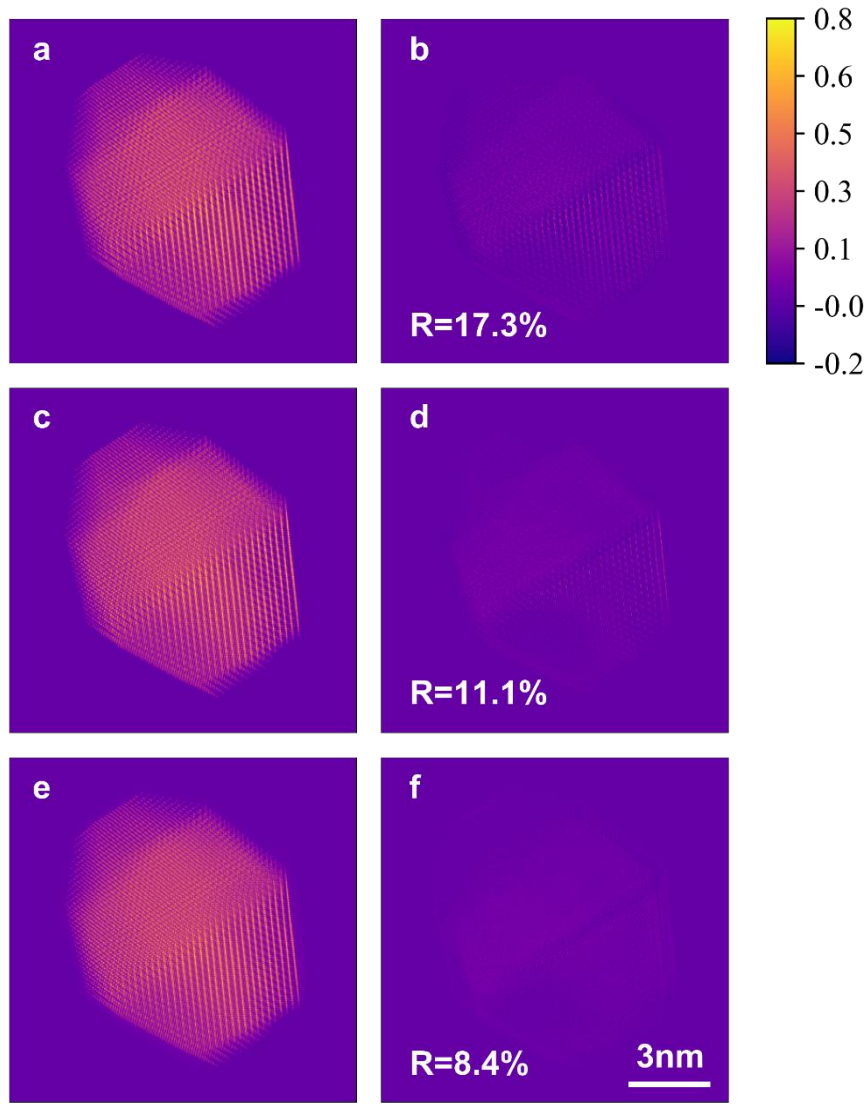

**Fig. 3. Comparison between tomographic reconstruction results with direct projection, rigid DDI, and adaptive DDI at a tilt angle of 0. a**. the projection image of the reconstruction with direct projection; **b**. the difference between a and the simulation image; **c**. the projection image of the reconstruction with rigid DDI; **d**. the difference between c and the simulation image; **e**. the projection image of the reconstruction with adaptive DDI; **f**. the difference between e and the simulation image. Scale bar, 3nm.

To obtain a more precise estimation of the quality of the reconstruction, we calculated the distances between the positions of the atoms in the reconstructed models and those in the original model (the ground truth). The distribution of the positional deviations obtained with the three probe functions is presented in **Fig. 4**. It is clear that the positional deviations obtained with direct projection are the largest. Both rigid and adaptive DDI consider the variation in depth of the interaction between the probe and the object. The reconstructions lead to narrower distributions of positional deviations.

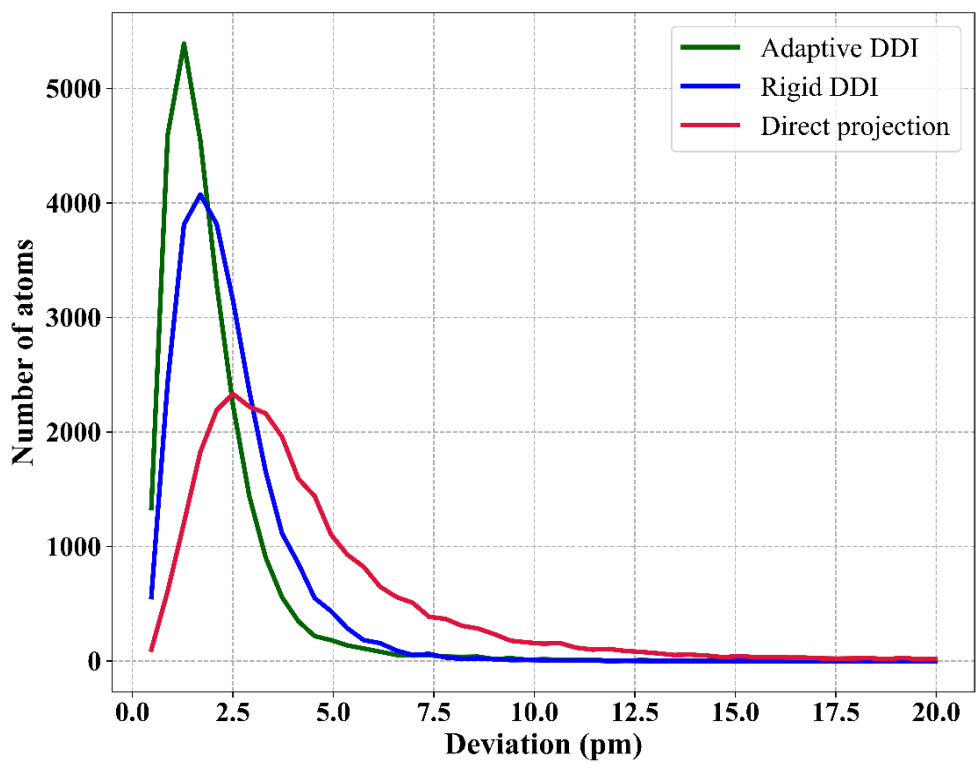

**Fig. 4. Deviation of the 3D atomic positions between the model of the $M_0$ and those reconstructed with direct projection, rigid DDI, and adaptive DDI.**

The root mean square deviations (RMSDs) between the atomic coordinates of the reconstructed model and the true coordinates are 4.3 pm, 2.2 pm, and 1.7 pm, for direct projection, rigid DDI, and adaptive DDI, respectively.

## LOT reconstruction on amorphous particles

To further test the LOT method for large-sized samples, we constructed particles composed of 25,000, 50,000, 75,000, 100,000, 150,000, and 200,000 silicon atoms. These were denoted as $M_1$ to $M_6$, respectively with 3D coordinates of the silicon atoms partially randomly distributed in space. The interatomic distances were constrained between 2.32 Å and 2.53 Å. The diameter of the six models were 9 nm, 12 nm, 14 nm, 15 nm, 17 nm, and 19 nm, for $M_1$, $M_2$, $M_3$, $M_4$, $M_5$, and $M_6$, respectively. Multislice simulations were performed using the same simulation parameters same as those for model

$M_0$. Typical simulation images are shown in **Fig. 5**.

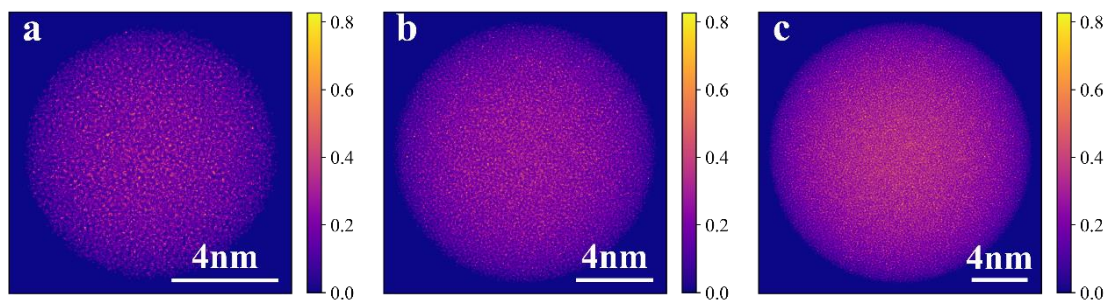

**Fig. 5. Multislice simulated HAADF-STEM images of three amorphous particles. a**, **b**, and **c** show the simulated images of model $M_1$, $M_3$, and $M_6$, respectively. Scale bar, 4 nm.

The RMSDs were used to evaluate the positional accuracy of the reconstructions for each model. As shown in **Fig. 6**, the positional accuracy decreases as the sample size decreases. The relationship between the number of atoms and the RMSD value is almost linear. The direct projection interaction type results in the lowest accuracy, with the positional deviation exceeding 20 pm for the 200k atom model. Both the rigid DDI and adaptive DDI significantly improve the positional accuracy. However, the adaptive DDI shows slightly better accuracy than the rigid DDI. This suggests that the probe-object interaction between the probe and object alters the electron beam as it travels through the object. Therefore, fine-tuning the probe function throughout the depth can further improve of reconstruction quality and positional accuracy.

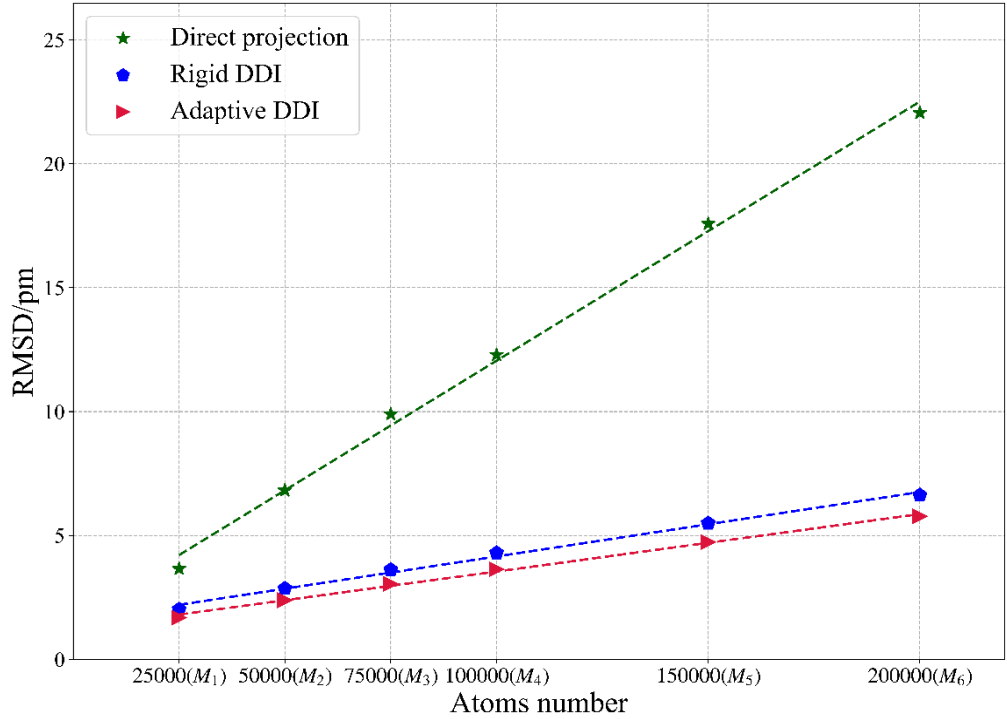

**Fig. 6. Positional accuracy of the reconstructed models using direct projection, rigid DDI, and adaptive DDI.**

## Conclusions

In summary, we have proposed and implemented the local-orbital tomography method. This method optimizes the optimization of the probe-object interactions to obtain accurate atomic positions. The results show that the LOT method yields high-quality reconstructions with precise atomic positions. Additionally, it can directly determine the chemical species of materials containing multiple elements. The LOT method is expected to have broad applications for atomic tomography, especially for large samples.

Real-time 3D analysis during electron tomography using tomviz, Nat Commun, 13 (2022) 4458.

# Methods

## Image Preprocessing

Various tilt series display variations in value distributions. To ensure consistent intensities of the atoms ($H_i$) during the reconstruction of each tilt series, we use a normalization technique that scales the tilt series within the range of 0 to 1:

$$I_j^{obs}(u,v) = \frac{I_j^{raw}(u,v) - min\left[I_j^{raw}(u,v)\right]_{j=1,2,\dots,M}}{max\left[I_j^{raw}(u,v)\right]_{j=1,2,\dots,M}} \tag{6}$$

where $I_j^{raw}(u,v)$ represents the $j$th raw measured image.

## Initialization of local orbitals and probe

The number of atoms and their coordinates are initialized as regular grids based on the estimated minimum atom spacing ($d_{est}$) of the sample. The parameters are given in **Supplementary Table 1**.

Based on the electron accelerating voltage, convergent beam half-angle, and layer thickness of the tilt series, we calculate the initial 2D probes $\{p_{jk}(u,v)\}_{j=1,2,\dots,M, k=1,2,\dots,n}$ corresponding to each angle and each layer depth.

The two-dimensional probe in the plane formed on the image plane of the condenser lens system can be expressed as:

$$\mathcal{F}\{I^{ADF}(r_p)\}(k_p) = 2C(\widetilde{R_N}, W) \cdot \overline{\mathcal{F}\{|\psi_{in}(r)|^2\}}(k_p) \cdot \mathcal{F}\{1 - cos\varphi(r)\}(k_p) \tag{7}$$

$$\psi_{in}(x,y) = \mathcal{F}_{2D}\{\psi(k_x,k_y)\}(x,y) \tag{8}$$

Here, $\psi(k_x,k_y)$ is the wave function of electrons at the back focal plane of the condenser system.

$$\psi(k_x,k_y) = A(k_x,k_y)e^{-i\chi(k_x,k_y)} \tag{9}$$

$A(k_x,k_y)$ and $\chi(k_x,k_y)$ are the aperture and aberration functions, respectively. $\mathcal{F}_{2D}$ is the two-dimensional Fourier transform operator.

In real space, a complete 3D probe can be obtained by propagating the two-dimensional plane electron beam forward and backward along the propagation direction z in free space. This is equivalent to directly applying the Fresnel propagation $e^{i\pi\lambda z\left(k_x'^2, k_y'^2\right)}$ operator to $\psi(k_x, k_y)$ in k-space.

Therefore, the 3D probe is:

$$\psi_{in}(r) = \mathcal{F}_{2D}\left\{\psi(k_x, k_y) e^{i\pi\lambda z\left(k_x'^2, k_y'^2\right)}\right\}(x, y) \tag{10}$$

Here, $r = (x, y, z)$ is now 3D. In other words, in the z-direction, we essentially calculate a defocused probe at a distance z from the image plane.

### Parameter optimization

After calculating the computation of images for each angle according to equation 1-5, the algorithm uses the gradient descent method to minimize the difference between the measured and calculated images. The gradient of the loss function with respect to each parameter to be optimized can be obtained using PyTorch's automatic differentiation library.

### Quality of reconstruction

We used the R-factor[11] to assess the difference between the reconstructed images and the simulation images:

$$R = \frac{\sum_{u=1}^{p}\sum_{v=1}^{q}\left|I_j^{obs}(u,v) - I_j^{cal}(u,v)\right|}{\sum_{u=1}^{p}\sum_{v=1}^{q}\left|I_j^{obs}(u,v)\right|} \tag{11}$$

### Acknowledgments

We thank H.Z. Sha and L.H. Liu for helpful discussions about the optimization of algorithms.

## Funding

This work was supported by Basic Science Center Project of the National Natural Science Foundation of China (52388201) and the National Natural Science Foundation of China (51525102).

## Author contributions

R.Y. conceived the idea and supervised the project. L.M. performed tomographic reconstruction and data analysis. J.C. performed image simulations. L.M. and R.Y. co-wrote the paper. All authors discussed the results and commented on the manuscript.

## Competing interests

The authors declare no competing interests.

## Data and materials availability

Correspondence and requests for materials should be addressed to Rong Yu.

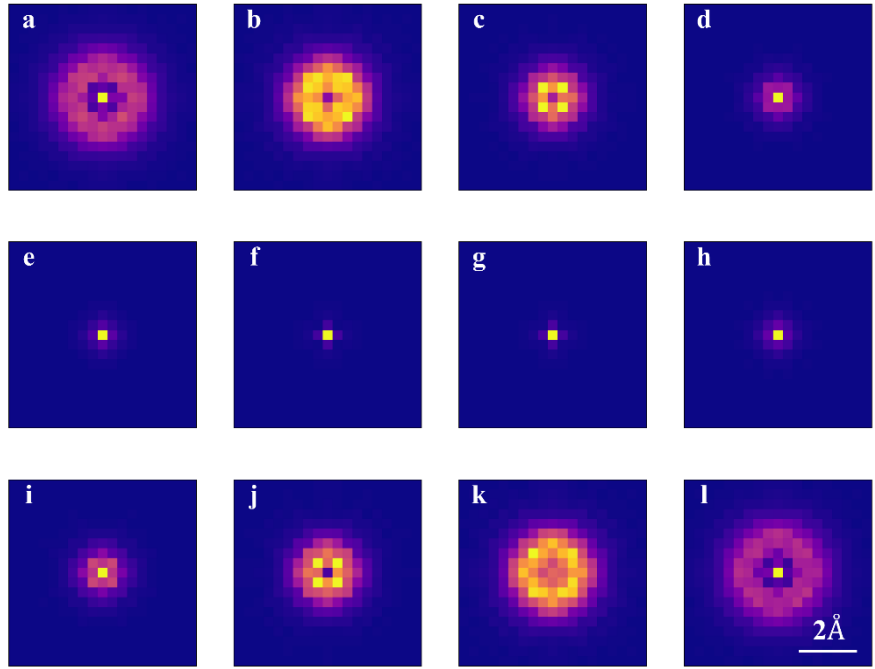

**Supplementary Figure 1 | Probe generated using simulation parameters for the model M$_6$ (without optimization). a** to **l** show electron beams at layers 3, 7, 11, 15, 19, 23, and 27, respectively, and the particle center corresponds to layer 17, with a slice thickness of 4 Å. Scale bar, 2 Å.

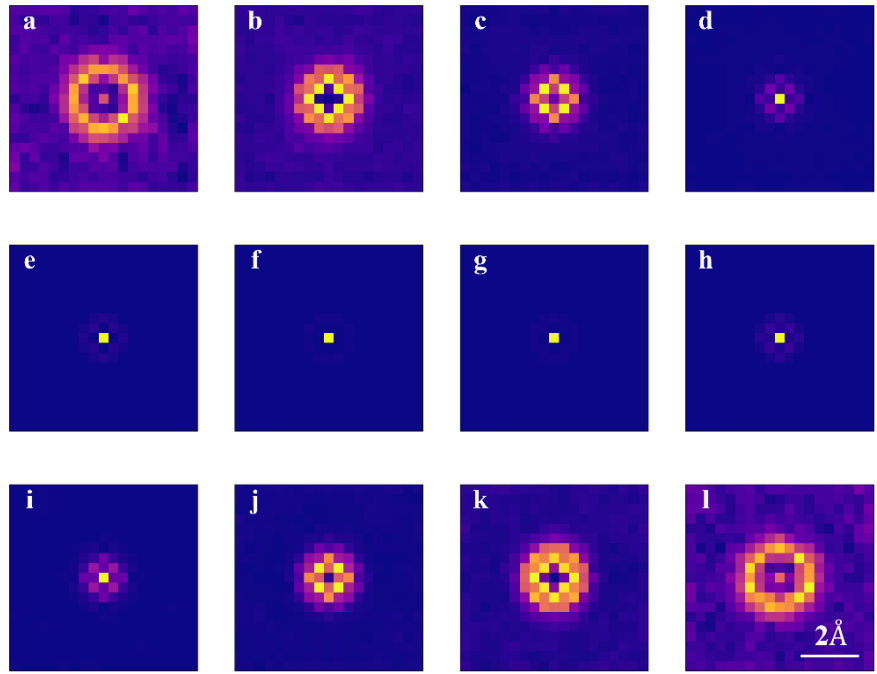

**Supplementary Figure 2 | Probe after pixelized optimization for the model M$_6$. a** to **l** show electron beams at layers 3, 7, 11, 15, 19, 23, and 27, respectively, and the particle center corresponds to layer 17, with a slice thickness of 4 Å. Scale bar, 2 Å.

**Supplementary Table 1| Initialization of parameters.**

| Parameter | Initial value |
|:---:|:---:|
| $H_i$ | 0.0080 |
| $B_i$ | 0.5 Å |
| $u_j$ | 0.0 Å |
| $v_j$ | 0.0 Å |
| $d_{est}$ | 1 Å |